\documentclass[sigconf]{acmart}
\usepackage{array}         % Column formatting
\usepackage{lineno}
\usepackage{tabularx}       % Flexible table widths
\usepackage{longtable}      % Tables that span multiple pages (if needed)

\usepackage{multirow}       % Cells spanning multiple rows
\usepackage{graphicx}       % For \rotatebox command (rotating text)

\usepackage{xcolor}         % Colors
\usepackage{colortbl}       % Colored table cells (\cellcolor)
\usepackage{enumitem}       % Custom itemize/enumerate formatting

\usepackage{booktabs}       % Professional looking tables (\toprule, \midrule, \bottomrule)
\usepackage{caption}        % Enhanced caption formatting

\usepackage{tikz}
\usetikzlibrary{positioning, arrows.meta, shapes}
\AtBeginDocument{%
  }

\copyrightyear{2025}
\acmYear{2025}
\setcopyright{rightsretained}
\acmConference[EAAMO '25]{Equity and Access in Algorithms, Mechanisms, and Optimization}{November 5--7, 2025}{Pittsburgh, PA, USA}
\acmBooktitle{Equity and Access in Algorithms, Mechanisms, and Optimization (EAAMO '25), November 5--7, 2025, Pittsburgh, PA, USA}
\acmDOI{10.1145/3757887.3763017}
\acmISBN{979-8-4007-2140-3/2025/11}

\begin{document}

%%
%% The "title" command has an optional parameter,
%% allowing the author to define a "short title" to be used in page headers.
\title{Anti-Regulatory AI: How ``AI Safety'' is Leveraged Against Regulatory Oversight}

%%
%% The "author" command and its associated commands are used to define
%% the authors and their affiliations.
%% Of note is the shared affiliation of the first two authors, and the
%% "authornote" and "authornotemark" commands
%% used to denote shared contribution to the research.
% \author{Anonymous Authors}
% \authornote{Both authors contributed equally to this research.}
% \email{trovato@corporation.com}
% \orcid{1234-5678-9012}
\author{Rui-Jie Yew}
\email{rui-jie_yew@brown.edu}
\affiliation{%
  \institution{Center for Technology Responsibility, Reimagination, and Redesign, Brown University}
  \institution{Center for Human-Compatible AI, UC Berkeley}
  \country{United States}
}

\author{Brian Judge}
\email{bjudge@berkeley.edu}
\affiliation{%
  \institution{Center for Human-Compatible AI, UC Berkeley}
  \country{United States}
  }

%%
%% By default, the full list of authors will be used in the page
%% headers. Often, this list is too long, and will overlap
%% other information printed in the page headers. This command allows
%% the author to define a more concise list
%% of authors' names for this purpose.
\renewcommand{\shortauthors}{Yew and Judge}

%%
%% The abstract is a short summary of the work to be presented in the
%% article.
\begin{abstract}
AI companies increasingly develop and deploy privacy-enhancing technologies, bias-constraining measures,  evaluation frameworks, and alignment techniques --- framing them as addressing concerns related to data privacy, algorithmic fairness, and AI safety. This paper examines the ulterior function of these technologies as mechanisms of legal influence. First, we examine how encryption, federated learning, and synthetic data --- presented as enhancing privacy and reducing bias --- can operate as mechanisms of avoidance with existing regulations in attempts to place data operations outside the scope of traditional regulatory frameworks. Second, we investigate how emerging AI safety practices including open-source model releases, evaluations, and alignment techniques can be used as mechanisms of change that direct regulatory focus towards industry-controlled voluntary standards and self-governance. We term this phenomenon \emph{anti-regulatory AI} --- the deployment of ostensibly protective technologies that simultaneously shapes the terms of regulatory oversight. Our analysis additionally reveals how technologies' anti-regulatory functions are enabled through framing that legitimizes their deployment while obscuring their use as regulatory workarounds. This paper closes with a discussion of policy implications that centers on the consideration of business incentives that drive AI development and the role of technical expertise in assessing whether these technologies fulfill their purported protections.
\end{abstract}

%%
%% The code below is generated by the tool at http://dl.acm.org/ccs.cfm.
%% Please copy and paste the code instead of the example below.
%%
\begin{CCSXML}
<ccs2012>
   <concept>
       <concept_id>10003456.10003462</concept_id>
       <concept_desc>Social and professional topics~Computing / technology policy</concept_desc>
       <concept_significance>500</concept_significance>
       </concept>
   <concept>
       <concept_id>10010147.10010178</concept_id>
       <concept_desc>Computing methodologies~Artificial intelligence</concept_desc>
       <concept_significance>500</concept_significance>
       </concept>
   <concept>
       <concept_id>10002978</concept_id>
       <concept_desc>Security and privacy</concept_desc>
       <concept_significance>300</concept_significance>
       </concept>
 </ccs2012>
\end{CCSXML}

\ccsdesc[500]{Social and professional topics~Computing / technology policy}
\ccsdesc[500]{Computing methodologies~Artificial intelligence}
\ccsdesc[300]{Security and privacy}
% %%
% %% Keywords. The author(s) should pick words that accurately describe
% %% the work being presented. Separate the keywords with commas.
\keywords{AI regulation, AI safety, AI alignment, privacy, security, fairness, legal avoidance, avoision}

% \received{20 February 2007}
% \received[revised]{12 March 2009}
% \received[accepted]{5 June 2009}

%%
%% This command processes the author and affiliation and title
%% information and builds the first part of the formatted document.
\maketitle
\section{Introduction}
As part of Anthropic's 2024 response to California's proposed SB1047, \citet{amodei2024sb1047} noted that the ``nascency'' of the science of AI evaluations ``is a reason not to legislate too prescriptively, too early.'' In response to a 2023 request for comment, OpenAI emphasized the importance of assessing AI safety and AI risks, suggesting that governmental efforts should be directed towards maturing currently ``nascent'' ``risk and capability evaluation practices''~\citep{openai2023request}. In a consultation documentation for the EU AI Act (AIA), Google noted that the use of privacy-enhancing technologies (PETs) should be prioritized in the development and deployment of AI systems, and that this priority would make it difficult to ``demonstrate compliance with dataset requirements'' under Article 10 of the draft or to ``provide direct access to datasets as per Article 64'' of the 2021 AIA draft.

In the 2020's, companies are wielding technical mechanisms from AI evaluations to PETs to shape the contents and effects of AI regulations. But this is not the first time that technical mechanisms have played such a role. In 2003, \citet{wu2003code} identified computer code's \emph{anti-regulatory} function as a ``tool to minimize the costs of law'' and as a ``mechanism of legal influence''. \citet{wu2003code} introduces this concept to describe the relationship between peer-to-peer networks (P2Ps) and copyright law, where the design and deployment of code minimized legal costs by circumventing or undermining legal constraints. By exploiting the technical architecture of these networks, P2P developers attempted to bypass the law that struggled to adapt to these new technological possibilities.  In elaborating his argument, \citet{wu2003code} identifies two anti-regulatory mechanisms: (1) \emph{mechanisms of avoidance} and (2) \emph{mechanisms of change}. Mechanisms of avoidance bypass costs of existing regulations or renders them harder to enforce, whereas mechanisms of change help secure future regulatory arrangements.

Using \citet{wu2003code}'s framework, \textbf{we catalog how technologies framed as enhancing AI safety, preserving privacy, and maintaining fairness can simultaneously function as anti-regulatory mechanisms of avoidance and change.} First, we illustrate how encryption, decentralization, and synthetic data --- technologies presented as privacy-preserving or bias-constraining --- can simultaneously operate as anti-regulatory \emph{mechanisms of avoidance} in attempts to place data operations outside of traditional regulatory conceptions. Second, we consider how the emerging pillars of the science of AI safety --- open-sourcing, evaluations/benchmarks, and alignment techniques --- can operate as anti-regulatory \emph{mechanisms of change}. By framing these technical toolkits as proactive measures for ensuring safety and reliability, companies redirect regulatory attention away from mandatory legal requirements and towards voluntary standards. This shift allows companies to preserve their autonomy while showcasing a commitment to responsible AI and the public interest.

For both mechanisms, we discuss (a) their rhetorical frames and (b) their anti-regulatory functions against the existing and emerging AI regulatory landscape. We do not argue that these technologies are necessarily ineffective at achieving their protective purposes, or that they are necessarily misaligned with their rhetorical frames. We aim to bring to light how evaluating these technologies through only their rhetorical frames can constrain how these technologies are conceptualized and, thus, the policy remedies that are considered. We conclude with a discussion of policy implications, emphasizing business incentives and the role of technical expertise in assessing technologies' anti-regulatory functions.
\section{Background}
\subsection{Anti-Regulatory Mechanisms}
On the cusp of the 21st century, \citet{lessig1999code} proclaimed that code is law. By constraining ``the very architecture of the particular [cyber]space''~\citep{lessig1999code}, computer code operated as law, and thus, as a kind of \emph{regulatory mechanism}~\citep{wu2003code}. But, in the early 2000s, technology played a central role in how compliance patterns shifted across domains from copyright to financial fraud. In response to code's role in the shifting compliance landscape, \citet{wu2003code} argues that code can also play an \emph{anti-regulatory} function, that code may not be a mechanism that in itself sets the law of cyberspace but one that responds to the law, that ``minimizes the cost of law''. \citet{wu2003code} identifies two primary avenues through which the costs of law can be minimized --- through \emph{avoidance mechanisms} and \emph{change mechanisms}. Avoidance mechanisms are ``efforts that would lower the expected costs of the law''. Change mechanisms ``(principally lobbying) decrease the sanction attached to certain conduct''~\citep{wu2003code}.\citet{wu2003code} devotes the rest of his analysis to applying this paradigm and motivating the function of code as a mechanism of avoidance, particularly in the context of peer-to-peer networks and copyright law. In this paper, we consider the role of technical mechanisms as both mechanisms of avoidance and as mechanisms of change.

\begin{table*}[htbp]
\centering
\caption{This paper uses the framework of anti-regulatory mechanisms put forth in~\citep{wu2003code} to reason about the role of code in AI's regulatory landscape. While \citet{wu2003code} focuses on the role of code primarily as an avoidance mechanism, we also consider the role of AI code as a mechanism of change.}
\label{tab:framework_comparison}
\begin{tabular}{p{2cm}p{3cm}p{4cm}p{4cm}}
\toprule
\textbf{Mechanism} & \textbf{Definition} & \textbf{Wu's Framework~\citep{wu2003code}} & \textbf{Anti-Regulatory AI} \\
\midrule

\textbf{Avoidance} & 
\begin{minipage}[t]{\linewidth}
Technology lowers regulatory costs by decreasing detection probability or enabling alternative conduct
\end{minipage} & 
\begin{minipage}[t]{\linewidth}
\textbf{Examples:}
\begin{itemize}[leftmargin=*,nosep]
\item Peer-to-peer networks (Napster, KaZaA)
\item File-sharing protocols
\end{itemize}
\end{minipage} & 
\begin{minipage}[t]{\linewidth}
\textbf{AI examples:}
\begin{itemize}[leftmargin=*,nosep]
\item Encryption
\item Federated Learning
\item Synthetic Data
\end{itemize}
\end{minipage} \\

\midrule

\textbf{Change} & 
\begin{minipage}[t]{\linewidth}
Mechanisms that influence regulatory development to secure favorable legal arrangements
\end{minipage} & 
\begin{minipage}[t]{\linewidth}
\textbf{Political processes:}
\begin{itemize}[leftmargin=*,nosep]
\item Lobbying campaigns
\item Litigation strategies
\end{itemize}
\end{minipage} & 
\begin{minipage}[t]{\linewidth}
\textbf{Technical mechanisms also as mechanisms of change:}
\begin{itemize}[leftmargin=*,nosep]
\item Open-Source Models
\item AI Evaluations \& Benchmarks  
\item AI Alignment Techniques
\end{itemize}
\end{minipage} \\

\bottomrule
\end{tabular}
\end{table*}

\subsection{Frames}
Reputation is a key consideration for corporate strategy: a good reputation helps companies remain competitive~\citep{wexler2012extralegal, roberts2002corporate, argenti2004reputation}. However, profit-maximizing practices, when exposed, can trigger concrete legal actions and damage reputational standing. These countervailing incentives --- pursuing profit while avoiding reputation damage --- can give rise to tactics like washing, offering pretextual arguments, and selective framing. Washing involves promoting a positive value as the defining quality of a practice, even when that practice produces negative effects~\citep{li2022clean}. Pretextual arguments leverage positive values or claim protection against negative outcomes as justification for certain conduct, while obscuring the true underlying motivations~\citep{van2022privacy}. These strategies create a pathway for corporations to simultaneously maximize profits and maintain positive public perception—effectively, having their cake and eating it too.\footnote{More specifically, as \citet{wexler2012extralegal} writes, washing enables actors to ``prevent the imposition of requested substantive changes and relatedly to avoid
reputational costs from litigation by appearing as being a good citizen on social issues''.} 

As part of our analysis, we discuss the largely corporate-created frame of AI technologies and their anti-regulatory functions. Defining an argument as pretextual can hinge on intentions or motivations. Defining washing can hinge on rhetoric that exaggerates or conceals a behavior's true effect. Used in a similar light to both washing and pretexts, framing can be more subtle than the use of either. \textbf{Framing has been defined as ``selecting and highlighting some facets of events of issues, and making connections among them so as to promote a particular interpretation, evaluation, and/or solution''~\citep{entman2009projections}.} By placing focus selectively, frames can play a significant role in shaping the discourse surrounding the function of a technology. For example, in the context of digital platforms, \citet{gillespie2010politics} considers the how companies use the term ``platform'' to frame the services they offer, and how this rhetoric is invoked in such a way as to elide tensions and avoid scrutiny.\footnote{``The term `platform' helps reveal how YouTube and others stage themselves for these constituencies, allowing them to make a broadly progressive sales pitch
while also eliding the tensions inherent in their service: between user-generated and
commercially produced content, between cultivating community and serving up advertising, between intervening in the delivery of content and remaining neutral''~\citep{gillespie2010politics}.} In some cases, the framing associated with a technology can  be  diversionary as regards to its effects in light of the regulatory landscape, but the technology itself may still confer significant protections. In others, however, the frame may, in fact, constitute a pretextual argument for the undesirable technological behaviors that underlie the framing. Our aim is to highlight the discrepancy between the technology's rhetorical framing and the role the technology can play in  regulatory avoidance and preemption.

\subsection{Related Work and Contributions}
A growing body of scholarship highlights the potential for firms to use technology to bypass governance attempts. ``Fairwashing'' has been identified as the practice of ``promoting the false perception that the learning models used by the company are fair while it might not be so''~\citep{aivodji2019fairwashing}.  Researchers have also identified certain corporate uses of privacy-enhancing technologies (PETs) to ``privacy wash''~\citep{shrishak2025privacy, dagstuhl25112}: to claim that a use of PETs protects privacy when that use in fact intrudes on user privacy or still has the potential to harm downstream users~\citep{leonte2025privacy}.\footnote{The California Lawyers Association has also warned against ``privacy-washing'', which ``occurs when a company advertises that it prioritizes data protection with its customer-facing products and services but neglects to actually implement best privacy practices to secure and minimize the processing of customers’ personal information''~\citep{jiminez2021privacy}.}  \citet{hooker2024limitations} similarly considers how techniques like federated learning can be used to get around compute thresholds as a form of AI governance. \citet{yew2025red} consider how firms might develop and deploy AI systems to avoid the EU AI Act (AIA). 

 There is also scholarship that addresses tactics surrounding the preemption of AI regulations. \citet{abdalla2021grey} likens the lobbying strategies employed by big tech to lobbying strategies employed by the tobacco industry. When talks of regulation escalated concurrently with a string of scandals in 2018, AI companies shifted discussions ``to focus on voluntary `ethical principles,' `responsible practices,' and technical adjustments or `safeguards' framed in terms of `bias' and `fairness' (e.g., requiring or encouraging police to adopt `unbiased' or `fair' facial recognition)''~\citep{ochigame2019invention}.  \citet{ochigame2019invention} argues that the invention of the emerging field of ``ethical AI'' was an attempt to shift regulatory focus away from the implementation of stringent rules and towards self-governance. By wielding technical~\citep{ochigame2019invention} and philosophical~\citep{bietti2020ethics} rigor, companies built legitimacy around the emerging field of ``ethical AI''. In Section~\ref{sec:change}, we consider a play from the same rhetorical playbook with the emerging science of AI safety. While \citet{ochigame2019invention} considers the rhetoric shaping the broader umbrella of the AI ethics toolkit, our analysis goes through specific technologies in the AI safety toolkit and consider how they and their rhetorical frame shapes AI's regulatory landscape. Similar to how we consider the ways in which AI safety technologies can be used to preempt regulatory efforts in Section~\ref{sec:change}, \citet{casper2025pitfalls} examines how ``evidence-based'' AI policy takes a page out of the ``deny and delay'' playbook used in debates surrounding tobacco and climate change, and can serve to delay regulation.

  In coupling our discussion of both a particular AI technology's rhetorical frame and its anti-regulatory function, we aim to demonstrate how a particular AI technology's anti-regulatory function often sits outside of its rhetorical frame. A technology's primary frame can play a critical role in how its functions are understood, and, thus, how it is governed. We aim to draw attention to the anti-regulatory role of these technologies beyond their rhetorical frames, including this role in the realm of policy considerations.

\begin{table*}[htbp]
\centering
\caption{Anti-Regulatory AI: Below is an overview of the avoidance and change mechanisms that we discuss in this paper.}
\label{tab:antiregulatory_mechanisms}
\renewcommand{\arraystretch}{1.4}
\begin{tabular}{p{4cm}p{4.5cm}p{6.8cm}}
\toprule
\textbf{AI Technology/Practice} & \textbf{Frame} & \textbf{Anti-Regulatory Function} \\
\midrule
\multicolumn{3}{c}{\textbf{Avoidance Mechanisms}} \\
\midrule
\textbf{Encryption} & Privacy, security, safety, sovereignty  & Use in attempts to move data outside the scope of laws that govern PII; avoids costs and reaps rewards of data sharing\\
\midrule
\textbf{Decentralized Technologies} & Privacy, collective benefit, power distribution & Use in attempts to bypass data protection laws through distributed processing; obfuscates data sourcing for copyright enforcement\\
\midrule
\textbf{Synthetic Data} & Bias reduction, privacy preservation, data diversity & Use in attempts to bypass consent and attribution requirements and avoid PII regulations \\
\midrule
\multicolumn{3}{c}{\textbf{Change Mechanisms}} \\
\midrule
\textbf{Open-Source Models} & Innovation, collective research advancement & Use in attempts to secure regulatory exemptions; leveraging of ambiguous definitions for competitive advantage; control of ecosystem development \\
\midrule
\textbf{AI Evaluations \& Benchmarks} & Scientific rigor, nascency & Promotes voluntary if-then commitments; positions industry as technical authority; legitimizes self-governance \\
\midrule
\textbf{AI Alignment Techniques} & Automatic, ``human-out-of-the-loop'' paradigm & Obscures labor dependencies; muddles liability attribution\\
\bottomrule
\end{tabular}
\end{table*}

\section{Mechanisms of Avoidance: Fair and Private Data Shelters for AI Training}\label{sec:avoidance}
\textbf{Mechanisms of avoidance ``decrease the probability of detection [of illegal behavior]'' or enable bypassing the classification of heavily regulated forms of conduct by ``adopt[ing] other forms of conduct with the same effect"}~\citep{wu2003code}. Intellectual property law has long seen the use of technological design decisions to minimize exposure to legal regimes. When deciding how to use an already-patented technology or discovery, companies weigh the costs of infringing, licensing, or ``designing around'' an existing patent claim—designing a technology with similar function to a patented one such that it does not touch on its protected aspects.\footnote{
``To be certain, competitors to a patent holder who invent around are avoiding the alternative of infringement, and they will presumably do so only if inventing around is cheaper than the alternatives of being penalized for infringing or licensing''~\citep{burk2014inventing}.} Designing around the copyright statute\footnote{\cite{burk2014inventing} describes that ``inventing around a patent involves skirting the definition of the protected invention, whereas inventing around a copyright involves skirting terminology in the copyright statute''.} has resulted in innovative changes to technologies. These designs exploited weaknesses in copyright enforcement and worked around the subject matter covered by its statutes. Just as monetary transactions can be structured to minimize taxes via innovative tax shelters, code can also be structured to place computations outside the scope of burdensome legal requirements~\citep{giblin2014aereo}.

Against the backdrop of long-standing regulations that implicate data --- vital for AI training and deployment --- similar logics of moving data towards less regulated categories may drive the use of PETs \footnote{The definition of PETs is contested, but they have been examined as a class of technologies that enable user control over their personal information,~\citep{tavani2001privacy} and technologies that reduce external exposure to user's personal information~\citep{kaaniche2020privacy}} and their building blocks. Generative AI, a popular class of AI tools that can produce ``high-quality artistic media for visual arts, concept art, music, and literature''~\citep{epstein2023art}, is a significant --- and data-hungry --- sector of the AI economy. OpenAI's GPT required more than 45TB of data from sources like Common Crawl, book corpuses, and Wikipedia~\cite{brown2020language}. The Pile dataset~\citep{gao2020pile}, used to train multi-modal models like Meta's Open Pre-trained Transformer~\citep{zhang2022opt} and allegedly used to train models released by NVIDIA, Apple, and Anthropic~\cite{gilbertson2024apple}, contains 825 GiB worth of pirated e-books and subtitles from YouTube videos. The value of data in the development of these AI technologies is not necessarily contained in any single data point and is extracted primarily through its large-scale accumulation~\citep{viljoen2021relational}.\footnote{As \citet{sadowski2019data} writes: ``[w]hen data is treated as a form of capital, the imperative to collect as much data, from as many sources, by any means possible intensifies existing practices of accumulation and leads to the creation of new ones.''}

We consider underlying business incentives in the use of encryption, decentralization, and synthetic data to avoid regulatory requirements along with the rhetorical framing that legitimizes their use. Whether the use of these techniques confer protections is orthogonal to our analysis. Rather, our focus is how, when wrapped in the rhetoric of privacy and fairness, the use of PETs navigate the trade-off between data accumulation and data liability where expedient. In particular, we focus on regulatory considerations along the axes of data protection\footnote{\citet{veale2023rights} describes data protection as the most salient legal regime touching user data.} and privacy laws, copyright law\footnote{There has been a significant amount of copyright litigation regarding the use of copyrighted works in the development and use  of AI technologies~\citep{samuelson2024think}.}, and data requirements as part of emerging AI policies. 

\subsubsection{Encryption}\label{sec:encryption}
Encryption obfuscates plaintext data, and is often used prevent unauthorized access to the underlying data.\footnote{For a more detailed overview surrounding encryption and encryption algorithms see, e.g., ~\citep{bhanot2015review}.} Encryption may also be end-to-end, as in end-to-end encryption (E2EE), wherein communication data is not only protected from third-parties but also from the primary data collector themselves~\cite{scheffler2023sok}.\footnote{Encryption can also be implemented with a focus on the prevention of third-party intrusion --- in which case, the data collector would have access to the underlying plaintext.} 

\paragraph{Frame}
``Safety'', ``security'', and ``privacy'' are highlighted as core values backing  Meta's decision to launch end-to-end by default on Facebook messenger ~\citep{crisan2023launching}.  E2EE is also highlighted under the header ``Building Privacy-First Products'' as part of a Meta blog post titled ``Investing in Privacy''~\citep{metaprivacy}. Privacy is highlighted as a primary rationale in the implementation of E2EE. Subsequent decisions made by other technology platforms to implement E2EE were also backed by similar values. Publicity surrounding Google's implementation of E2EE on its Android messaging services as well as the company's implementation of E2EE for enterprise emails emphasize security, confidential communication, and sovereignty~\citep{google2024end, googlee2eeenterprise}. Beyond its use in the messaging communications context, encryption is more broadly framed as a security and privacy enhancing technology, from its use on the cloud to its use in a wide range of data-sharing applications. In a blog post detailing advancements in homomorphic encryption, a technique that enables training machine learning models on encrypted data, Amazon Web Services (AWS) describes the technology providing ``an extra layer of security''~\citep{amazonmlencryption}.

Given the growing global priority of AI sovereignty~\citep{WEF2024SovereignAI}, encryption has also been framed as a technology of sovereignty. It is included as part of AWS's sovereign-by-design principles, with the header ``[t]he ability to encrypt everything everywhere''~\citep{aws_digital_sovereignty}, as part of Microsoft's sovereign cloud offerings~\citep{microsoft2025sovereignty}, and even as a part of press release for Gmail's end-to-end encryption for enterprises service~\citep{google_gmail_e2ee_2024}.

\paragraph{Avoidance}
Under costly regulatory regimes, there are also concrete business incentives that surround how encryption is developed and deployed. With E2EE, communication services stand to save costs resulting from third-party requests for internal communications --- such as those from regulators or law enforcement. The lack of access to plaintext message data can allow companies to side-step entirely costly ``quandaries''~\citep{veale2023denied} about information access and moderation, such as government subpoenas requesting message data. Encryption as it relates to sovereignty technologies might also allow companies to play both sides --- convincing entities who use their services that their data is secure and sovereign even as regulations like the Clarifying Lawful Overseas Use of Data could require cloud companies to hand over that data to the U.S. government~\citep{woollacott2025microsoft}.

Deployments of encryption also play an important role in the accumulation of data against data protection and privacy regimes. Across approaches to data protection and privacy around the globe, these laws are triggered through interfacing with personal and sensitive data~\citep{solove2023data} --- typically discussed as personally identifying or identifiable data (``PII''). These protections heighten with the sensitivity of data~\citep{solove2023data}: as the degree to which data is identifying or identifiable increases, so too do the requirements that accompany them.

When data is encrypted, it has been argued that these laws no longer apply to resulting data operations. This argument routes through rhetoric that the data is no longer PII or personal data. For instance, Google has argued that its ``double-blind encryption technology [] prevents both Google and our partners from viewing our respective users' personally identifiable information''~\cite{mark_bergen_google_2018}. PII has also been defined as ``a specific category of particularly sensitive data defined as \emph{unencrypted} electronic information''. 

In 2018, MasterCard and Google cut a deal to share encrypted data in order to track retail sales~\citep{mark_bergen_google_2018}. As part of this deal, private set intersection (PSI), was used to combine user data about physical retail sales with data about their online activity. PSI leverages encrypted information to enable the confirmation of whether an ad seen online resulted in a purchase in-stores. This technology created a pathway for Google to consolidate even more granular data about its users and to monitor and influence consumers. These models could then be trained with encrypted data in a manner aligned with Google's business objectives: to influence consumers and generate revenue for its advertising partners. Throughout these transactions, companies upheld the framing of privacy and security through announcements that encryption protected users' plaintext personally identifiable information~\cite{mark_bergen_google_2018}.

Finally, encryption enables companies to extract value from data sharing while maintaining competitive boundaries. Encryption allows companies to selectively share specific insights without revealing the underlying datasets that comprise their core business.\footnote{Managing third-party access through encryption has been recognized as a competitive strategy. \citet{jones2020nonrivalry} discusses how encryption can be used to exclude access from competitors --- increasing the value of data to the company with unimpeded access to it.}.  The Google-MasterCard collaboration illustrates this dynamic: rather than exchanging raw datasets, the companies used encryption (specifically PSI) to share only the minimal information necessary for their mutual benefit --- confirming whether online ads led to in-store purchases. This selective sharing mechanism allowed both companies to collaborate on valuable analytics while protecting their respective data assets from full exposure to their partner. Through encryption, firms can thus participate in data-sharing arrangements that would otherwise be too risky from a competitive standpoint, expanding the ways data can generate value without sacrificing exclusive control.

\subsubsection{Decentralization}
In traditional AI development, data consolidation and model training occurs ``under the authority of a single entity''~\citep{shadmy2024reimagining}. This means that data is accessible by and computed on a centralized server. Techniques for decentralization such as federated learning (FL) and multi-party computation (MPC) allow for data consolidation and model training to occur in a decentralized fashion. With FL, intermediate computations, such as model weight updates, are executed locally on user devices and then the individual results are aggregated for model training~\citep{zhang2021survey}. MPC allows for data sharing and computations between multiple entities, where data's underlying plaintext never needs to be revealed. This technique enables entities to collaborate without revealing their data to each other.

\paragraph{Frame} Decentralization promises the development of AI down a path of data sharing and collective benefit\footnote{In the original paper detailing the technology, \citet{mcmahan2017communication} details that federated learning presents an opportunity to ``collectively reap to the benefits of shared models trained from this rich data, without the need to centrally store it''. Additionally, \citet{shokri2015privacy} notes the competition benefits of decentralized AI training.}, privacy protection\footnote{For example, \citet{kumar2024enable, shadmy2024reimagining} highlight how federated learning (FL), a decentralization technique, addresses security and privacy concerns.}, and the decentralization of power.\footnote{\citet{shadmy2024reimagining} discusses ML decentralization's ``emancipatory appeal'' and how the story of decentralization embeds a promise of restructuring power.}
Decentralization has enabled collaborations between hospitals and medical facilities~\cite{sheller2020federated}, as well as between government agencies~\cite{treiber2022data,kamara2021decentralized}.  These use cases are front-and-center as part of promotional materials for decentralization AI technologies. With the header ``Federated Learning to save the day (and save lives)'', an Amazon Web Services blogpost recommends datasets that can be used on their platform~\citep{kumar2024enable}.
Use cases for medical diagnoses and healthcare improvements are highlighted as part of NVIDIA's promotional material for the adoption of its federated learning (FL) --- a decentralization technique --- platform, FLARE~\citep{dogra2021federated}. The use of FL on Microsoft Azure has also been promoted as helping doctors with medical transcript analyses~\citep{chambers2022microsoft}.

\paragraph{Avoidance}
Like uses of encryption discussed in Section~\ref{sec:encryption}, decentralized training techniques for AI can be employed to avoid the purview of data protection laws. \citet{shadmy2024reimagining} write that: ``[a] central incentive that motivates and drives the development and adoption of federated learning is the possibility of producing models from personal data which is problematic to collect directly due to ethical, legal, or organizational constraints''. Under data protection laws, data is often conceptualized as being collected, possessed, or processed. Traditionally, data computations are conceptualized as being done by a data processor or a data controller. Within environments for decentralized data consolidation and AI training, users may themselves contribute to computing towards infrastructure that serve to profile them and others like them ~\citep{veale2023rights}. With decentralized AI training, there is then an argument that data is not interfaced in a way that triggers the application of data protection laws. 
 
 Technologies for decentralized AI training might also be used to avoid copyright liability. FL in combination with differential privacy has the widely-publicized effect of conferring protection to the information contained in data, but it can also serve another purpose ---  to obfuscate where and how data is sourced. It can serve to erect barriers between data and the firms that collect it in the name of privacy. Indeed, as part of a 2021 response to an AIA proposal consultation, Google pushed for a prioritization of the PETs over AIA dataset requirements. Google argued that, because the use of PETs like FL for decentralization would prevent the company from directly interfacing with datasets, it would not be possible for certain AI system to ``demonstrate compliance with dataset requirements'' under Article 10 of the draft or to ``provide direct access to datasets as per Article 64'' of the draft.  This could also position the company outside the scope of copyright enforcement and redress~\citep{google2021consultation}.\footnote{While the use of PETs is not explicitly exempt from the final draft of the AIA, the requirements in the AIA to provide summaries of datasets are conditioned upon companies' business interests and trade secrets~\citep{eu2024aiact}.} 

\subsubsection{Synthetic Data}
Synthetic data is defined as: ``artificial data, generally generated by computer simulations or algorithms, which has analytical value''~\cite{gal2023synthetic, rubin1993statistical}. For AI development, synthetic training data is often generated by AI models themselves.\footnote{AI language models such as NVIDIA's open-source Nemotron-4 340B~\citep{adler2024nemotron} and Cosmopedia released on HuggingFace~\citep{benallal2024cosmopedia} generate synthetic text data. Synthetic image datasets have also been produced through models like Stable Diffusion~\citep{tian2024stablerep}.}, and synthetic datasets have been used to train models that detect aircraft in satellite imagery~\cite{shermeyer2021rareplanes} and recognize faces~\cite{qiu2021synface}. There is a body of scholarship that investigates settings under which synthetic data helps or harms model learning. In particular, model collapse has been discussed as a potential detrimental consequence of training on AI-generated synthetic data~\citep{shumailov2024ai, shumailov2023curse}. Research has also suggested that ``accumulating'' synthetic data and training models with real data in addition to synthetic data can in fact boost model performance across a variety of settings~\citep{ kazdan2024accumulating, gerstgrasser2024model}.

\paragraph{Frame}
Synthetic data is positioned as a solution to data scarcity while promising to mitigate bias in AI systems ~\citep{riemann2024synthetic, zhao2024taxonomy}. For instance, IBM Research explains: ``[s]ynthetic data is information that's been generated on a computer to augment or replace real data to improve AI models, protect sensitive data, and mitigate bias''~\cite{martineau2023synthetic}. Synthetic data has also been praised for its privacy-preserving potential --- described as having ``the potential to improve privacy and representation in artificial intelligence''~\citep{savage2023synthetic}. Similarly, in a Microsoft Research blogpost, \citet{afonja2024crossroads} promotes the role of synthetic data as a way to address privacy concerns while maintaining important innovation advantages. Synthetic data has also been cited within The House Task Force on Artificial Intelligence Final Report as offering ``privacy-enhancing benefits, given that it does not include information on real individuals''~\citep{taskforce2024bipartisan}. 

\paragraph{Avoidance}
The rapidly growing appetite for data in AI development inevitably conflicts with regulations like copyright and data protection laws. Requirements for consent\footnote{Consent is a requirement as part of ~\citep[Art. 6/7]{eugdpr}, as well as across multiple approaches to regulating sensitive data within data protection regulations~\citep{solove2023data}.} and attribution\footnote{Attribution is key component of copyright law ~\citep{copyrightlaw}.} pose costly obstacles. Synthetic data offers a convenient escape from these constraints ~\citep{whitney2024real}.  Training AI models on synthetic data has been described as a form of ``copyright laundering''~\cite{newton-rex2024tweet} because AI outputs are not considered copyrightable in the U.S. as of 2024. As part of a 2023 lawsuit which concerns the copyrightability of AI outputs, the U.S. Copyright Office ruled that ``human authorship is an essential part of a valid copyright claim'' and denied copyright protections for the generated AI output in question~\citep{crs2023generative}. The Copyright Office's decision is currently under appeal, but this could mean that synthetic data could then be used to make an argument that it does not have a human author as conceptualized in copyright law --- and thus has no copyright protections. Adobe has touted the fact that its Firefly image generation model is designed to generate content ``that does not infringe on copyright and other intellectual property right[s]''~\cite{smith2024growing}. Yet as much as 14\% of its training data may consist of images generated by another AI model (Midjourney) that was itself trained on real-world data~\cite{metz2024adobe}. It has been said that synthetic data is being used in this manner more broadly across the AI industry~\cite{newton-rex2024companies}. 

There are also two ways in which synthetic data can function as an avoidance mechanism under data protection and privacy laws. Fist, synthetic data may not be collected or processed in the sense that triggers data protection laws but is, instead, arguably ``generated''~\cite{gal2023synthetic}. This provides an argument that synthetic data may not be regulated under existing legal approaches to data protection and privacy. Secondly, there is an argument that synthetic data falls outside of data covered under the scope of data protection and privacy laws. As \citet{gal2023synthetic} note:``both the sectoral focus of many privacy laws and the way in which they define the term `personally identifiable information' (`PII') imply that even their application to collected data may be limited...the capacity of these laws to capture synthetic data, which is one step further removed from the individual, is more doubtful''. O'Reilly's textbook for using synthetic data in AI further details: ``Given that synthetic data would not be considered identifiable personal data, privacy regulations would not apply, and obligations of additional consent to use the data for secondary purposes would not be required''~\cite{el2020accelerating}.

\section{Mechanisms of Change: Regulatory Backburning with the Emerging Science of AI Safety}\label{sec:change}
Industries from telecommunications to chicken farming have combated current and potential regulation with influence campaigns. This strategy is built on the recognition that "information and how it is interpreted are integral to how government makes decisions on an issue''~\citep{teachout2014market}. The industry playbook follows a predictable pattern: when facing potential regulation, companies invest heavily to shape expert discourse in their favor. Through strategic alliances between industry groups, non-profits, and academics, they sponsor research that provides intellectual legitimacy for cheaper alternatives.\footnote{For example, in the wake of the 2007 financial crash, the Securities Industry and Financial Markets Association paid professors at elite institutions to comment on proposed interventions, with most of the comments opposing the changes. When the Food and Drug Administration (FDA) began to put out rules that would place closer scrutiny on the power imbalances in the chicken farming industry, alliances like the National Chicken Council and the National Meat Association funded research that concluded that the proposed rules would cost the industry over \$1 billion dollars over five years~\citep{teachout2014market}.}  The AI industry has rapidly adopted these established influence strategies, with the number of companies participating in AI lobbying more than doubling between 2022 and 2024~\citep{merica2024ai}. These lobbying efforts employ multiple mechanisms to shape the policy discourse and landscape~\citep{wei2024ai, wsj_silicon_valley_ai_pacs_2025}. Companies have also positioned themselves as essential partners in AI education, conducting technical workshops and establishing collaborations and commercial relationships with academic institutions~\citep{meta2024llm}.

This section examines how technical design choices and safety measures have the strategic effect of reducing regulatory impact. Leading AI companies have developed comprehensive internal policies and evaluation frameworks ostensibly designed to manage safety risks from advanced AI models. However, these self-regulatory measures can serve a dual purpose: they project an image of responsible governance while preempting external regulatory oversight. Like in backburning\footnote{Thanks to Stephen Casper for this term.}, where an uncontrolled fire is fought with controlled fire, the otherwise shape of external oversight becomes constrained to industry self-governance through the strategic use of AI safety measures.

\subsubsection{Open-Source}
AI models can have varying \emph{modes of deployment}, i.e., they can be released with varying degrees of openness. However, the definition of open-source remains vague. \citet{liesenfeld2024rethinking} write that ``openness is not a binary feature'' and \citet{solaiman2023gradient} posits that open source releases exist on a gradient of openness. On one end of the taxonomy of open-source models presented  in~\citep{solaiman2023gradient}, AI models might be deployed in a ``fully closed'' fashion, meaning that the AI system is inaccessible to those outside of the firm and used for internal research only. On the other end, AI models can be deployed in a ``fully open'' fashion, where, not only is the model completely accessible, but the datasets are also publicly released and available. There are also deployment modes that lie somewhere in the middle. For instance, a model can be accessed through an API, where the provider retains a degree of control over what is accessible through the API. Or, the model may be accessible to end-users through a web interface.

\paragraph{Frame}
Within the software industry, open-source has historical connotations in innovation~\citep{kogut2001open}. Open-source AI has leaned into the innovation connotations of open-source software. For example, the rippling, innovative effects of open-sourcing Unix as Linux is detailed within the open-source manifesto published on the Facebook blog~\cite{zuckerberg2024opensource}. Google has similarly brought the relationship between open-source and innovation to the fore: ``...OSS is critically important to AI innovation''~\citep{google2021consultation}. ``[O]pen-source AI drives responsible innovation'', argues an IBM-sponsored piece in The Atlantic~\citep{ibm2024open}. Companies including Meta, IBM, and Oracle comprise AI Alliance, which is ``committed to support and enhance open innovation across the AI technology landscape''~\citep{aiallianceabout}. The opportunity for increased collective research output is commonly cited as a reason to open-source AI models~\cite{leahy2021release}. 

\paragraph{Change}
The rich history of open-source and software innovation has been leveraged to lend weight to the notion that open-source AI is an endeavor that deserves protection from regulatory scrutiny.  As part of Google's response to an early draft of the AIA, the company argues that AIA should ``[m]ake clear that open-source software is out of scope'' because ``OSS [open-source software] is critically important to AI innovation''~\citep{google2021consultation}. This clear exemption was ultimately granted as part of the resulting AIA.\footnote{``AI systems released under free and open-source licences, unless they are placed on the market or put into service as high-risk AI systems or as an AI system that falls under Article 5 or 50''~\citep[Art. 2(12)]{eu2024aiact}.} In an effort to change potentially incoming AI policy in California, the AI Alliance, a coalition of companies, non-profits, and government organizations, advocated for ``incorporating discrete open-source exceptions throughout the bill [California's SB 1047] to ensure that responsible developers can open-source models with adequate mitigations''~\citep{aialliance2024statement}. These ``discrete open-source exceptions throughout the bill'' would exempt regulated entities' open models from many of the bill's original requirements.

Ambiguity regarding exact specifications of open-source in both existing and upcoming regulations presents an opportunity for AI firms to put forth specifications of open-source that are self-serving~\citep{widder2024open}. The terms ``open'' and ``open-source'' have been used in ``confusing and diverse ways'' and more as ``aspiration'' or ``marketing'' than a ``technical descriptor''~\citep{widder2023open}. The rhetoric of open-source that is put forth by technology companies has impact in how technologies can fall into less burdensome legal classifications.\footnote{As \citet{pollman2019tech} writes: ``The tech industry is notorious for design that pushes the regulatory envelope and aggressively uses rhetoric to defy legal norms and shape legal classifications.''} Rhetorical efforts by the AI industry are already taking shape. After Meta released its Llama language models as ``open-source'' despite dispute as to whether the released models are actually ``open-source'', several media outlets regurgitated the company's language of ``open-source''~\cite{fried2024meta} --- perpetuating rhetoric which could allow Meta to fall into the less burdensome category of open-source while not ensuring the transparency that accompanies the historical connotation of open-source.

\subsubsection{AI Evaluations}
The evaluation of models on benchmark datasets is a primary currency of research progress within the field ~\citep{raji2021ai} --- they set the standards for measuring progress in AI. AI production and research communities have propped up benchmark \emph{leaderboards} as important markers of performance and progress~\citep{mazumder2024dataperf}. \citet{ethayarajh2020utility} warns of using evaluation benchmarks as a proxy for progress, but they also attribute the rapid progress in  natural language processing (NLP) to the ``leaderboard paradigm'' surrounding evaluation benchmarks like GLUE~\citep{wang2018glue} \footnote{Examples include the ``Measuring massive multitask language understanding'' benchmark~\citep{hendrycks2020measuring}, the ``A Graduate-Level Google-Proof Q\&A Benchmark''~\citep{rein2023gpqa}, and ARC.} Evaluations have also been developed to measure ``the extent to which models are fair~\citep{tamkin2023evaluating},reliable~\citep{perez2022discovering}, honest~\citep{lin2021truthfulqa}, or less prone to malicious use~\citep{li2024wmdp}''~\citep{ren2024safetywashingaisafetybenchmarks}. As part of an ``AI safety benchmark''~\citep{vidgen2024introducing}, that is ``designed to assess the safety risks of AI systems that use chat-tune Language  Models''.  \citet{xu2022safebench} provide a benchmarking  platform to measure the safety of autonomous vehicles. In practice, audit tools are an important way to evaluate AI systems, but there are also drawbacks regarding how these tools fulfill evaluation needs along the axes of information access, regulatory guidance, and technical standards ~\citep{ojewale2024aiaccountabilityinfrastructuregaps}. 

\paragraph{Frame} 
Corporate rhetoric surrounding evaluations is deeply tied to voluntary \emph{if-then commitments}~\citep{karnofsky2024if}. If-then commitments usually come in the following form: if the AI is evaluated to reach a particular capability level, then a set of corresponding safety measures will be enacted~\citep{karnofsky2024if}. Such commitments are found in corporate AI safety frameworks like Anthropic's ``Responsible Scaling Policy'' (RSP), which sets four ``AI Safety Levels'' (ASL) modeled on biosafety levels~\citep{anthropic2024responsible},  OpenAI's ``Preparedness Framework''~\citep{openai2023preparedness}, and DeepMind's ``Frontier Safety Framework'' (FSF)~\citep{deepmind2023frontier}. As part of these frameworks, evaluations are invoked to assess everything from ``misaligned AI'' to ``chemical, radiological, and nuclear risk''~\citep{deepmind2023frontier}, to cybersecurity risks~\citep{anthropic2024responsible}. 

As part of these frameworks, evaluations are framed as a rigorous way to measure the capability of AI models in relation to the different capability tiers. For \citet{anthropic2024responsible}, evaluations are described as providing ``strong evidence'' regarding the skill of a model. \citet{openai2023preparedness} describes OpenAI's investment in developing and adopting evaluations that are ``science-backed'' and that provide ``high precision and high recall indications of whether a covered system has reached a capability threshold in one of our Tracked Categories''. 

The rigor and scientific nature of AI evaluations is further buttressed by third-party evaluators and governmental AI safety and security institutes, who partner with AI corporations~\citep{meyer2025ukaisafety, nist2025aisic}. The UK Secretary of State for Science, Innovation and Technology describes the work of the UK AI Safety Institute (AISI)   as science-driven, noting that the work of the AI Safety Institute ``will move the discussion forward from the speculative and philosophical, further towards the scientific and empirical''~\citep{ukgov2024aisafety}. The United States AISI is housed within the National Institute of Standards and Technology (NIST) --- known for its risk management framework and approaches across a range of disciplines --- and pursues projects like ``measurement science for AI safety'', ``safety evaluations of models and systems'', and ``guidelines for evaluations and risk mitigations''~\citep{nist2024aisi}.
 
\paragraph{Change}

The frames of science, rigor, and nascency surrounding AI evaluations have been invoked in a feedback loop within voluntary if-then commitments and AI evaluations in the preemption of more binding regulations. The \emph{science} of AI evaluations is \emph{rigorous}: we can depend on evaluations to tell us whether the models have reached a particular capability threshold. However, the science of AI evaluations is itself \emph{nascent}, meaning that we must continue invest in research surrounding AI evaluations. These research efforts then legitimize AI evaluations as rigorous, feeding back into supporting if-then voluntary commitments as a sufficient means of AI regulation. \citet{ochigame2019invention} categorizes the most desirable regulatory regimes for technology companies as, firstly, those that leave the implementation of ``responsible practices'' as ``merely voluntary'' and, secondly, those that ``encourag[e] or requir[e] technical adjustments that do not conflict significantly with
profits''. In this section, we show how the industry-backed governmental and academic ecosystem bolstering the ``nascency'' of the science of AI safety evaluations prioritizes voluntary, technical, and non-binding regulation. By focusing on a nascent science and its iterations, companies can remain operations in these most desirable buckets of voluntary and non-binding regulations.

The ``nascency'' of AI evaluations and safety has been used to motivate the need to iterate on safety frameworks --- and to ultimately to advocate for deregulation or hands-off regulation. With the rhetoric that ``the science of AI risk evaluations is nascent''~\citep{openai2023request}, OpenAI emphasizes the need for a ``better science of AI risks'' ~\citep{openai2023request} and the importance of government efforts towards maturing ``risk and capability evaluation practices''. As part of Anthropic's response to California's SB1047, \citet{amodei2024sb1047} claims that the nascency of this science `` is a reason not to legislate too prescriptively, too early''~\citep{amodei2024sb1047} and motivates that there are governmental  `` benefits of pushing AI companies to invest in developing this science [of AI risk reduction]''. 

Companies like OpenAI and Anthropic have stood behind the development of AI safety institutes and AI security institutes, providing monetary support to the UK AISI~\citep{perrigo2025uk}. Crucially, the UK AISI ``often does not publicly disclose the results of its evaluations, nor information about whether AI companies have acted upon what it has found, for what it says are security and intellectual-property reasons''~\citep{perrigo2025uk}. The role of government in the form of safety and security institutes can then become transmuted from ensuring safety to convincing the public that this privatized safety regime is sufficient to make AI safe trustworthy. On the academic front, the AI Safety Fund --- funded in part by  Google, Microsoft, Anthropic, and OpenAI --- has awarded its first round of grants for technical research by ``independent researchers,'' on topics like ``[e]valuating and assessing strategies for addressing [safety-critical risks posed by frontier models]''~\citep{fmf2024aisafety}. Meta has similarly awarded its ``Large Language Model (LLM) Evaluation Research Grant'' to fund ``university faculty''  to conduct research that ``generates novel, challenging, ground truth benchmarks and evaluations, both for pre-training and post-training''~\citep{meta2024llm}.

Just as the telecommunications industry and meat packing industry funded influence campaigns and academic research to legitimize their policy positions~\citep{teachout2014market}, the legitimacy of the science of evaluations bolstered by independent, corporate, and academic researchers can serve as a backdrop to the preemption of AI regulations. The policy priority then becomes developing this ``nascent science''. By creating a self-fulfilling ecosystem of industry, academic, and government research under the header of ``third-party evaluations''~\citep{anthropic2024third}, companies create a rigorous regulatory front  of governance surrounding evaluations, but one that may not pursue meaningful evaluations that expose drawbacks of their AI models and systems to scrutiny.\footnote{Such a regime is also reminiscent of the relationship between Wall Street and government-approved credit rating agencies, where credit rating agencies had incentives to provide positive ratings. Because credit agencies made money through the banks that hired them to rate their bonds, these agencies had every incentive to provide banks with good ratings.} This regime is one that legitimizes safety frameworks, which, when deployed, may not do much to change current AI practices and services --- Anthropic states explicitly that ``from a business perspective, we want to be clear that our RSP will not alter current uses of Claude or disrupt availability of our products''~\citep{anthropic2023rsp}.

\subsubsection{Alignment}
The discussion of alignment in a technical context typically relates to reinforcement learning and, under the current paradigm of AI development, reinforcement learning from human feedback (RLHF). With reinforcement learning, a reward function is specified by the AI developer and, with a good algorithm, the AI system  performs well according to this function~\citep{hadfield2019incomplete}. 
However, optimizing for the specified function presents opportunities for misalignment, which is defined as a mismatch between an AI's specifications and the behavior of the AI~\citep{zhuang2020consequences}. Misaligned AI can manifest as deceptive behavior~\citep{hubinger2024sleeper} and dishonest responses~\citep{bang2023multitask,ji2023ai}. On the other hand, the alignment of AI, AI steered toward human values~\citep{ji2023ai}, has been described as being able to be achieved through techniques like reinforcement learning with human feedback~\citep{casper2023open} in the pre-deployment stage or through discriminative assessment~\citep{mazeika2025utility, khan2025randomness} and generative approaches~\citep{mazeika2025utility, khan2025randomness} in the post-deployment stage.

\paragraph{Frame}
The ability of AI to align themselves or to guide themselves to safety is a characteristic highlighted by corporate state-of-the-art AI alignment techniques. \citet{openai2025improving} discusses ``rule-based rewards'' (RBRs) as ``significantly enhanc[ing] the safety of our AI systems''. Unlike RLHF, which requires human feedback, to enhance model behavior, RBRs does not require ``recurrent human input'' to evaluate model outputs. As part of Constitutional AI, AIs themselves are ``enlisted to supervise other AIs'' and identify harmful outputs without any human involvement~\cite{bai2022constitutional}. The capabilities of AI systems to evaluate other AI systems are sometimes even painted as surpassing the capabilities of humans to do the same. For instance, \citet{perez2022red} notes: ``Overall, our results suggest that some of the most powerful tools for improving LM  safety are LMs themselves''.

\paragraph{Change}
While it is rarely disputed that the use of AI to align models grants efficiency improvements, a focus on the ``human out the loop'' as part of alignment techniques can serve to detract from human labor involved in the development and deployment of AI in the first place. As \citet{perrigo2023exclusive} writes of the lack of disclosure surrounding how data and evaluation  work is contracted and outsourced: ``This may be the result of efforts to hide AI’s dependence on this large labor force when celebrating the efficiency gains of technology. Out of sight is also out of mind.''

By drawing regulatory attention toward high-level AI capabilities and behaviors and away from the labor force that it depends on, a ``human-out-the-loop'' alignment paradigm has the potential to preempt human governance efforts. It can serve to support a rhetoric that AI alignment and red-teaming efforts are an activity that does not warrant human intervention and scrutiny. This is already playing out, with the Archive Unit of Parliament in the EU noting in documents that Constitutional AI ``ensures'' that AI follows constitutional rules~\citep{iccl2025deploy}. In fact, the language of ``compliance'' has been deployed by the EU Parliament to describe Constitutional AI, that the AI is designed in ``compliance'' with a Constitutional AI approach~\citep{iccl2025deploy}. This language frames the AI as as being inherently compliant, standing in for human oversight. Human-out-the-loop alignment also has concrete consequences for the attribution of liability. Where a particular entity or person would have been responsible for safety violations involving AI systems, the automation of safety or testing pipelines can muddy attribution of liability. \citet{nissenbaum1996accountability} points to how computing and automation can serve to create ``vacuums of accountability''. Where there once would have been an entity to point to in the attribution of liability, the potential absence of an entity to point to or the ``blaming of the computer'' creates an avenue for indemnification.

\section{Conclusion}
Techniques nominally deployed to enhance AI safety and trustworthiness can serve dual purposes: while presented as prudent, proactive measures, they can simultaneously or alternatively function as ``anti-regulatory" mechanisms, influencing the shape of compliance with current and future regulations. Policy initiatives emphasizing science-based AI governance highlight the critical need to understand both AI's internal mechanisms and its associated risks~\citep{bommasani2024path}. However, our work demonstrates that attention must also be paid to the incentive structures that shape AI production. When regulatory discourse fixates on AI's technical complexities --- its opacity, unpredictability, and uncontrollability --- it can inadvertently elevate technical measures as universal solutions. This narrow focus then obscures crucial questions about whether these technical deployments genuinely provide the protections they promise or merely serve corporate interests. Only through focusing regulatory attention on the structures that \emph{surround} the black box's construction~\citep{selbst2024deconstructing} --- business incentives and even regulatory exposure --- can questions regarding the actual protections of technical deployments and their alignment with regulatory values be properly assessed. While addressing the anti-regulatory dimensions of AI technologies requires looking beyond purely technical features or approaches for governance, our call does not diminish the need for technological expertise, which remains important in evaluating industry claims. We must maintain critical awareness that, when technical solutions are promoted as remedies for AI-related harms, these measures may be strategically co-opted to serve business interests rather than public protection. Understanding this dynamic is essential for developing regulations that achieve their intended protective goals rather than merely satisfying procedural requirements.

\begin{acks}
We thank Suresh Venkatasubramanian, Stephen Casper, Kris Shrishak, Bill Marino, Lucy Qin, colleagues at UC Berkeley and Brown, and anonymous reviewers for their helpful insight.
\end{acks}
\bibliographystyle{ACM-Reference-Format}
\bibliography{sample-base}
\end{document}